\documentclass[sigconf, authorversion]{acmart}
\usepackage{balance}
\AtBeginDocument{%
 }

\copyrightyear{2025}
\acmYear{2025}
\setcopyright{rightsretained}
\acmConference[SIGCSE TS 2025]{Proceedings of the 56th ACM Technical Symposium on Computer Science Education V. 1}{February 26-March 1, 2025}{Pittsburgh, PA, USA}
\acmBooktitle{Proceedings of the 56th ACM Technical Symposium on Computer Science Education V. 1 (SIGCSE TS 2025), February 26-March 1, 2025, Pittsburgh, PA, USA}
   
\acmDOI{10.1145/3641554.3701880} \acmISBN{979-8-4007-0531-1/25/02}

\makeatletter
\gdef\@copyrightpermission{
  \begin{minipage}{0.3\columnwidth}
   \href{https://creativecommons.org/licenses/by/4.0/}{\includegraphics[width=0.90\textwidth]{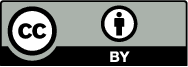}}
  \end{minipage}\hfill
  \begin{minipage}{0.7\columnwidth}
   \href{https://creativecommons.org/licenses/by/4.0/}{This work is licensed under a Creative Commons Attribution International 4.0 License.}
  \end{minipage}
  \vspace{5pt}
}
\makeatother

\begin{document}


\title{Teaching Program Decomposition in CS1: A Conceptual Framework for Improved Code Quality}


\author{Georgiana Haldeman}
\orcid{0000-0001-6046-5924}
\affiliation{%
  \institution{Colgate University}
  \city{Hamilton, NY}
  \country{USA}}
\email{ghaldeman@colgate.edu}

\author{Judah Robbins Bernal}
\orcid{0009-0003-8924-561X}
\affiliation{%
  \institution{Colgate University}
  \city{Hamilton, NY}
  \country{USA}}
\email{jrobbinsbernal@colgate.edu}

\author{Alec Wydra}
\orcid{0009-0008-3013-9974}
\affiliation{%
  \institution{Colgate University}
  \city{Hamilton, NY}
  \country{USA}}
\email{awydra@colgate.edu}

\author{Paul Denny}
\orcid{0000-0002-5150-9806}
\affiliation{%
  \institution{University of Auckland}
  \city{Auckland}
  \country{New Zealand}}
\email{paul@cs.auckland.ac.nz}

\renewcommand{\shortauthors}{Georgiana Haldeman, Judah Robbins Bernal, Alec Wydra, and Paul Denny}

\begin{abstract}
    Program decomposition is essential for developing maintainable and efficient software, yet it remains a challenging skill to teach and learn in introductory programming courses. What does program decomposition for procedural CS1 programs entail? How can CS1 students improve the decomposition of their programs? What scaffolded exercises can instructors use to teach program decomposition skills? We aim to answer all these questions by presenting a conceptual framework that (1) is grounded in the established code style principles, (2) provides a systematic approach that can be taught to students as an actionable strategy to improve the program decomposition of their programs, and (3) includes scaffolded exercises to be used in classroom activities. In addition, this systematic approach is automatable and can further be used to implement visualizers, automated feedback generators and digital tutors. 
\end{abstract}





\maketitle

\section{Introduction}

% Software quality is such an important and challenging topic among professionals and researchers that its introduction in early programming courses has been emphasized 
Introducing software quality in early programming courses has been emphasized due to its importance and difficulty and to ensure that students gain the necessary knowledge and experience for success in advanced courses and later in their careers \cite{teachingSoftwareQuality}. Specifically, introducing the topic of structural quality of code early is viewed as essential in laying the groundwork for more advanced treatment in later courses \cite{stegeman2014towards}. Also, there is strong motivation among instructors to review novices’ code more systematically because low-quality code can be a sign of students' misconceptions or poor design skills \cite{stegeman2014towards}. 

The structural quality of code could be viewed as something to be reviewed and assessed. There are many principles, design patterns, metrics and tools that are aimed at supporting the manual review of the structural quality of code or assessing and correcting it automatically \cite{stegeman2014towards,marinescu2001detecting,yin2024lizard}. However, the tools and approaches used by both professionals and researchers are not easily adaptable to the needs of educators and their students. For example, many of these tools are developed for the object-oriented paradigm (OOP) \cite{marinescu2001detecting,rizwan2020theoretical}, while many CS1 courses are taught with a focus on the procedural paradigm (PP) \cite{vilner2007fundamental, mason2024global}; often, the transfer of these principles across paradigms is not straightforward, even for instructors.

Structural quality of code could be also viewed as the result of the design process. There are two approaches that stand at the ends of the design process continuum: top-down and bottom-up. The top-down approach starts with the larger problem and repeatedly breaks it down into its component parts. The bottom-up approach starts with component parts and repeatedly merges them to solve the larger problem ~\cite{bottomUpTopDown, topDownDifficulties}. Many common design guidelines tend to be top-down~\cite{castro2015investigating}, which work well for experts, but may pose additional challenges for novices ~\cite{topDownDifficulties}. 

% As part of conceptualizing this framework, we explore the transfer of software design principles and metrics from the predominant OOP to the PP targeted by CS1 courses~\cite{mason2024global}. 

% corresponds to an optimal program decomposition which can be used as a visual pedagogical tool

In this paper, we propose a conceptual framework to teach the structural quality of PP code. As part of our conceptual framework, we outline a set of decomposition-oriented exercises, provide bottom-up guidelines for novices, and present a coloring procedure of the data dependency graph (DDG) of programs that can be used to visualize an optimal program decomposition.  A resource containing 13 procedural decomposition exercises for use by educators is also provided as an online appendix\footnote{The resource can be accessed at: \href{https://doi.org/10.5281/zenodo.12797876}{https://doi.org/10.5281/zenodo.12797876} }. 

%Taken together, in this work we present a comprehensive procedural decomposition curricula component for CS1.

%Under our conceptual framework, we outline a set of decomposition oriented assignments, bottom-up guidelines for novices, and an automated method of evaluating decomposition in CS1 assignments that we validate under Kane's framework of test evaluation \cite{kaneTestEvaluation}. Taken together, we present a comprehensive procedural decomposition curricula component for CS1.

\section{Related Work}

Similar to other aspects of code quality, decomposition is an important and challenging topic that is often neglected in introductory courses. Although we have yet to fully understand its relation to code comprehension~\cite{tempero2024comprehensibility}, decomposition is a component of computational thinking~\cite{hsu2018learn}. Moreover, based on interviews with CS educators, ~\cite{kallia2021threshold} has concluded that procedural decomposition is a threshold skill. This means that procedural decomposition is a skill that requires both higher order thinking and practice to master, which "unlocks" many other concepts and skills for a student afterwards. After studying the perspectives of students, educators, and developers on code quality, ~\cite{borstler2017perceptions} concluded that code quality should be discussed more thoroughly in educational programs. The teaching needs encompass: defining program decomposition in the context of the introductory courses, strategies for novices, and scaffolded exercises and assessments. 

% analytically assessing and actively improving the decomposition of programs

\paragraph{Assessment} 
Multiple techniques of assessing code quality, including program decomposition, have been proposed out of which some are manual ~\cite{stegeman2014towards, kirk2024literature}, automated ~\cite{keen2015program} or hybrid \cite{chren2022evaluating}. The automated techniques are based on code metrics, such as, cyclomatic complexity~\cite{briand1999unified} most commonly. However, code metrics~\cite{briand1999unified,mijavcreusability} can be paradigm specific ~\cite{chidamber1994metrics} and often fail to robustly quantify the code quality of programs~\cite{fenton2014software} because of their summative nature that abstracts away many important details about the code. Several efforts have been made to validate these code quality assessment rubrics using the contents of several books on code quality combined with norms used by teachers of introductory programming courses ~\cite{stegeman2014towards} or software product quality models from the literature together with feedback from practitioners ~\cite{kirk2024literature}. Nonetheless, neither provides actionable strategies or scaffolded exercises, which is the gap we aim to fill with our conceptual framework. 

Assessing code quality can be viewed as a process that the student engages with as well as an end product that can be evaluated. Charitsis et al. ~\cite{charitsis2023detecting} has introduced an automated approach for evaluating how, when and why students decompose code and compared results with development speed and exam performance. In this study, we also manually evaluated the reasons why programs in a CS1 curriculum should be decomposed and our findings complement those in ~\cite{charitsis2023detecting}.

\paragraph{Practices, Exercises and Tools} Stegeman et al. ~\cite{stegeman2016designing} has researched the designing of rubrics for code quality that is aimed at providing formative feedback to students. They stated that formative feedback is most effective when \textit{the learner has to (a) possess a concept of the standard (or goal, or reference level) being aimed for, (b) compare the actual (or current) level of performance with the standard, and (c) engage in appropriate action which leads to some closure of the gap} ~\cite{sadler1989formative}. Actionable techniques are consistent with the methods of supporting working memory~\cite{smith2018using}. Charitsis et al.~\cite{charitsis2022using} and Pearce et al.~\cite{pearce2015improving} have presented interventions for improving program decomposition skills of students. {\v{R}}echt{\'a}{\v{c}}kov{\'a} et al.~\cite{vrechtavckova2024catalog} compiled a catalog of code quality issues for introductory courses, but there are virtually no examples for program decomposition in it, a gap that this paper also addresses.

One active area of research in code quality education is the development of resources and tools to aid students in refactoring \cite{codeQualityEducationReview2023, izu2022resource}. Jiang et al.~\cite{jiang2020comparecfg} employed visualizing control flow graphs to provide visual feedback on code quality. Another particularly fruitful technique for program decomposition is program slicing. Program slicing is done using static analysis to identify cohesive subsections of a function that are related by a shared dependency \cite{kanemitsu2011visualization,ardalani2023supporting}.
The use of slicing in software refactoring is a well-established practice that has also been applied in an educational context for teaching novices \cite{teachingProgrammingSlicing2016, garg2018earthworm}. Our framework builds on this prior work by directly associating program slicing techniques with a comprehensible set of design principles that can be used for automated assessment and feedback generation.

\section{Proposed Conceptual Framework}

We propose a conceptual framework for the program decomposition of CS1-level procedural paradigm (PP) programs that is:

\hspace{1ex} \textbf{Systematic: }it provides a step-by-step procedure to reason about the decomposition of CS1-level programs that can be taught to students. Code quality can be viewed as a process as much as an end product. Our proposed conceptual framework was designed with the specific aim of aiding students in the decomposing process.
 
\hspace{1ex} \textbf{Rooted in well-established general principles: } Code quality topics such as decomposition go beyond the scope of CS1 to more advanced courses taught in different programming languages and paradigms and into the professional environment~\cite{borstler2017perceptions}. Thus, the learning transfer needs of code quality are both near and far~\cite{perkins1992transfer}. To stimulate the far transfer it's important to deliberately make connections across courses and professional settings. For that reason, our conceptual framework maps to well-established principles of code quality.
 
\hspace{1ex} \textbf{Automatable: } The systematic approach can be automated using static analysis similarly to the technique employed by the automated extract method refactoring by Ardalani et al.~\cite{ardalani2023supporting}. Given the increasingly widespread interest in CS education, automatable solutions are very attractive on the small and large scale for practice and revision of the course material, automated assessments, and feedback generation~\cite{haldeman2021csf}.

\subsection{Reasons for adding a function in CS1}

Broadly, program decomposition aids with program comprehension by leveraging the cognitive scaffolds provided by abstractions. In our context, the only method of abstracting is the adding of a function. Although abstracting code to a function is not inherently optimal~\cite{tempero2024comprehensibility}, it's often an effective means of improving code structure. Through our examination of CS1 assignments at a small liberal arts college (SLAC), we identified several reasons for adding a function:

%we identified the following relevant contexts:

%While adding a function is not always optimal as we previously discussed, there are multiple reasons why adding a function may make sense.  By going though all the programs in the CS1 curriculum at a small liberal arts college, we identified several reasons for adding a function:

\begin{enumerate}
    \item To split up the overall functionality of the program in concrete problem-specific sub-tasks. In the majority of cases---especially at the CS1-level---the optimal program decomposition trivially maps to certain specifications of the problem. For example, in the Garden problem (see Figure \ref{fig:garden} and further detail in Section \ref{sec:conceptual_framework}), three of the functions implement a different specification requirement of the problem, namely, the amount of plants, fill and soil separately. 
    \item To remove identical or functional duplication and increase reusability (potentially, with different inputs). These code snippets generally have a linear relationship between them which can be leveraged to collapse them into one function. For example, in the Garden problem, the computation of \texttt{circle\_plants}  is double the computation for \texttt{semi\_plants}.
    \item To abstract computationally complex code statements, for example, mathematical formulas such as the circle area formula. 
    \item To abstract programming-specific tasks, for example, transforming data from one data type to another, for instance, from a string to an array of decimals; data types and casting from one data type to another is an aspect that is specific to computer science and programming and has no relation to the problem domain. 
    \item To facilitate automated testing. This reason may be more advanced and therefore beyond the scope of CS1. 
\end{enumerate}

Our conceptual framework focuses on reasons 1-4 since the fifth category is more closely related to testability and maintainability~\cite{borstler2017perceptions}, which are more advanced code style concerns that are generally not addressed at the CS1 level.

\subsection{Overlapping Code Quality Principles}
\label{subsec:PDPrinciples}

We identified the following principles as central to the definition of procedural decomposition:

\hspace{1ex} \textbf{The single responsibility principle}, which is labeled as functional cohesion by Stegeman et al.~\cite{stegeman2014towards}, states that each function must have one well-defined purpose. It ensures that each function is cohesive, making the resulting code more readable and reusable.

\hspace{1ex} \textbf{Reduction of artificial coupling} is a principle mentioned in Martin's \textit{Clean Code}~\cite{martin2009clean} concerning limiting the amount of information that is passed around a program's structures unnecessarily. Code with high levels of coupling between functions is difficult to maintain as when future functionality is added, the relationships between coupled methods must be maintained. 

\hspace{1ex} \textbf{Reusability} refers to the degree to which functions defined in a program are likely to be interwoven into future processes \cite{mijavcreusability}. Reusability in code often makes the program easier to understand than the alternative of entire repeated code segments. As a result, decomposing code into functions with high levels of reusability and implementing reuse allows for code that is easier to understand, maintain, and extend.

\subsection{Conceptual Framework}
\label{sec:conceptual_framework}

\begin{figure}
    \centering
    \includegraphics[width=0.5\textwidth]{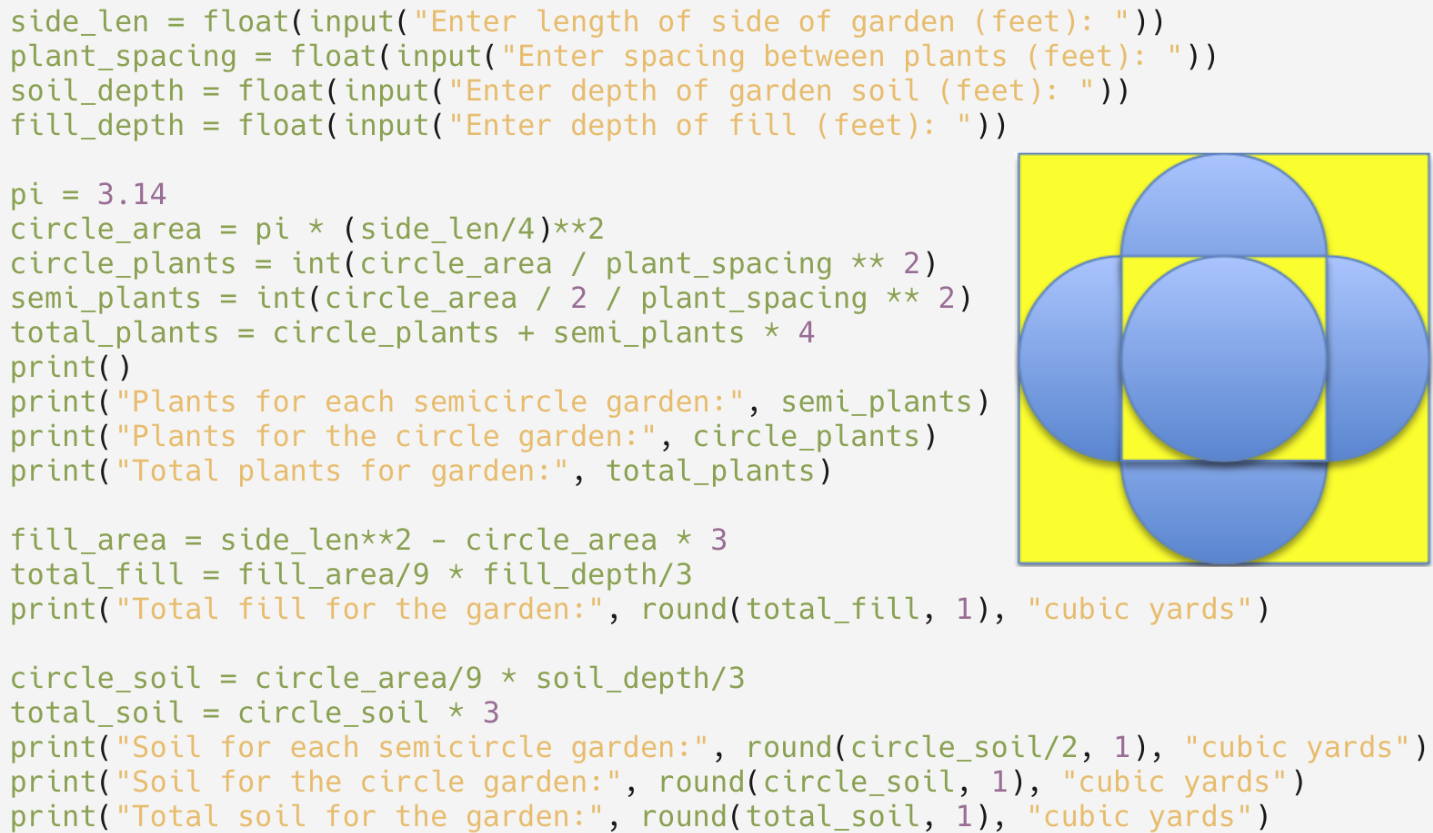}
    \caption{Global code and layout for the Garden problem.}
    \label{fig:garden}
\end{figure}

We define optimizing structural quality of code in the context of procedural paradigm (PP) programs at the CS1 level as the balancing of three main concerns that are consistent with the empirically validated model by Stegman et al.~\cite{stegeman2014towards} and the literature-informed model by Kirk et al.~\cite{kirk2024literature}: \textbf{(1) removing duplication and increasing reusability, (2) adhering to the single-responsibility principle and (3) reducing information passing between the independent functions of the program}. Concerns (1) and (2) target the functional similarity and dissimilarity (or low cohesion) of code, while (3) targets the data dependency across different components of the program which can be depicted using data dependency graphs (Section~\ref{subsec:visualizations}). To illustrate this, we refer to the problem called \textit{Garden} that requires students to calculate and print the amount of plants, fill and soil needed in different parts of a garden with a geometric layout (Figure~\ref{fig:garden}). For this problem, students are provided with a program containing a set of global code statements which are functionally correct but not decomposed and are required to decompose them across different functions (Figure~\ref{fig:garden}).

\begin{figure}
    \centering
    \includegraphics[width=0.5\textwidth]{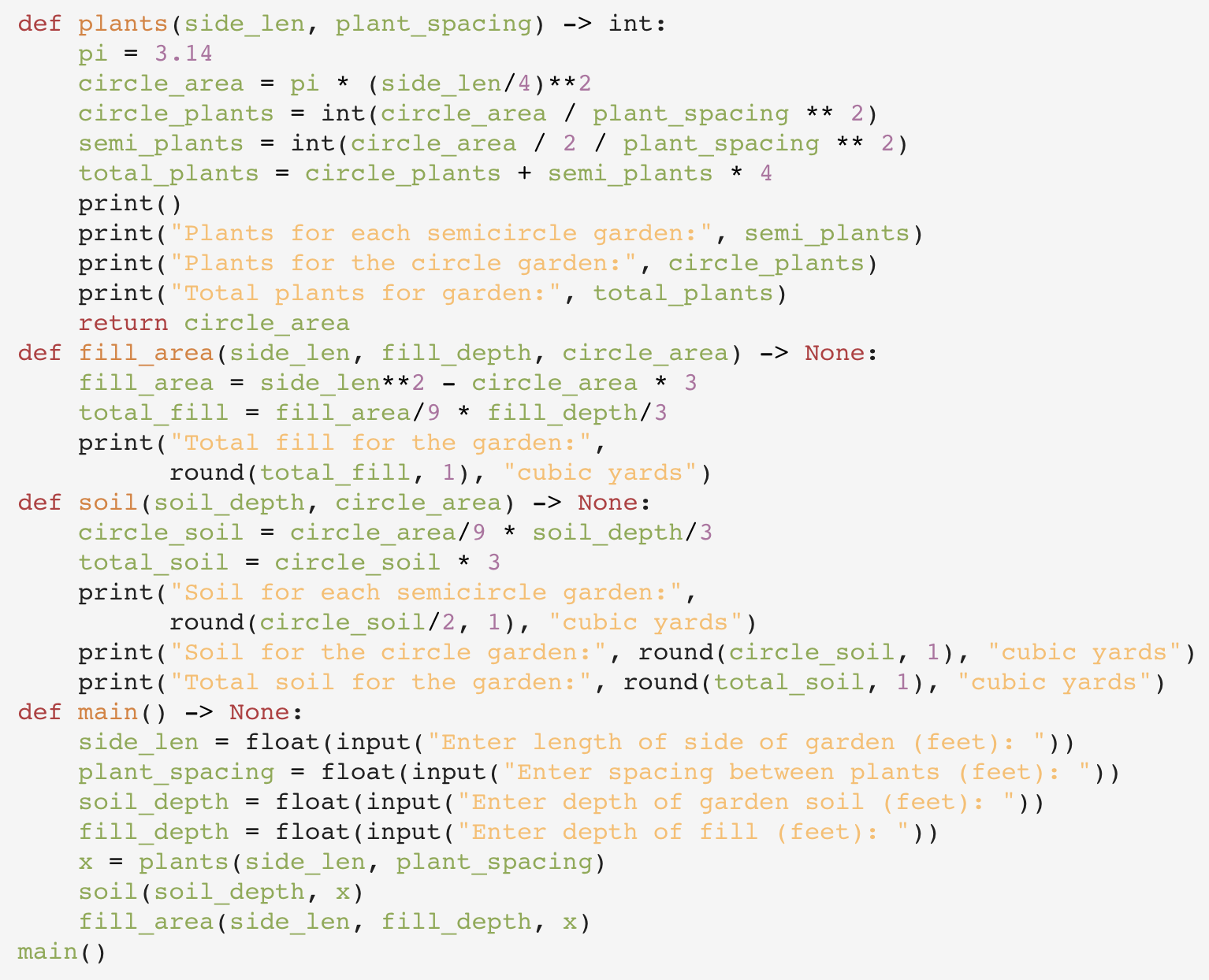}
    \caption{Student decomposition that violates the single responsibility principle.}
    \label{fig:garden_ex1}
\end{figure}

\begin{figure}
    \centering
    \includegraphics[width=0.5\textwidth]{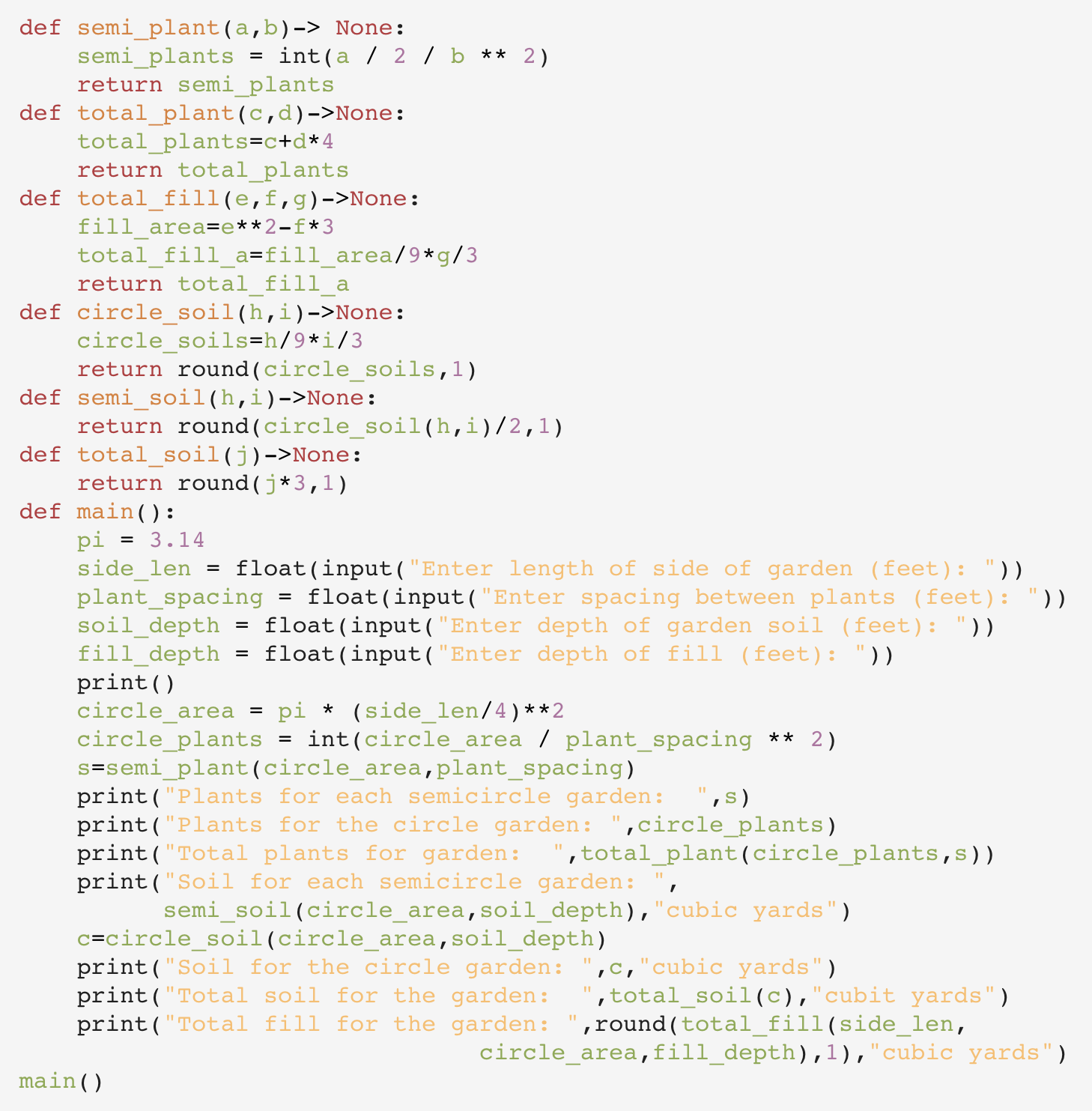}
    \caption{Student decomposition that violates the reducing information passing principle as well as the single responsibility principle.}
    \label{fig:garden_ex2}
\end{figure}

We define \textit{optimal procedural design} as a design that optimizes the structural quality of the code. Note that it is often the case that there are multiple optimal procedural designs for a given code. Furthermore, there are many more suboptimal procedural designs that students can easily arrive at without necessarily following a rigorous method. We show two representative samples collected from students in Figure~\ref{fig:garden_ex1} and ~\ref{fig:garden_ex2}. The design in Figure~\ref{fig:garden_ex1} reduces the overall information passing needed between the different parts of the program but violates the single responsibility principle because the \texttt{plants} function computes and returns the area of a circular section in the garden as well as computing and printing the plants needed. The design in Figure~\ref{fig:garden_ex2} increases the single responsibility measures at the cost of increasing the overall information passing between the different parts of the program by splitting up the code into more granular functions.

% \begin{figure}
% \includegraphics[width=.4\linewidth]{images/garden.png}
% \caption{The shape of the garden in the Garden Problem}
% \label{fig:garden}
% \end{figure}

% \begin{figure}
%     \centering
%     \begin{subfigure}[b]{0.4\linewidth}
%         \centering
%         \includegraphics[width=\textwidth]{images/garden.png}
%         \caption[]%
%         {{\small Garden Layout}}    
%         \label{fig:garden}
%     \end{subfigure}
%     \hfill
%     \begin{subfigure}[b]{0.5\linewidth}  
%         \centering 
%         \includegraphics[width=\textwidth]{images/owls.png}
%         \caption[]%
%         {{\small Exercise requiring drawing two owls (identical shape) separated by a random distance}} 
%         \label{fig:owls}
%     \end{subfigure}
%     \caption{Different layouts used by the proposed exercises.} 
%     \label{fig:exercises}
% \end{figure}

\subsection{Actionable Strategy for Novices}
\label{subsec:novices_procedure}

Up to this point, we've primarily discussed program decomposition as something to be assessed. However, improving program decomposition is also something that we do. We therefore present an actionable strategy for novices to improve the program decomposition of their CS1 procedural paradigm programs. Note that although our step-by-step procedure assumes that the students start with a program that is not decomposed, the procedure can be easily adapted to partially or poorly decomposed states as we further discuss in Section~\ref{sec:case_study}. In addition, observe that this procedure assumes that the code provided is trivially decomposable, which is not always the case. For example, code with an interleaved pattern composition~\cite{ginat2009interleaved} is generally harder to decompose and thus it is generally avoided in CS1 curriculum. The steps of the procedure are as follows:

\begin{enumerate}
    \item First, isolate parts of the code that depend on each other in terms of information (that is, the information stored by a variable is computed using the information from another variable). High cohesion between variables is generally indicative that the code statements work towards a common task. Wrap these high cohesion statements with function headers and give them appropriate names. 
    \item Next, refine the function extraction by collapsing functions with identical or functional duplication and isolating the programming-specific tasks such as transforming data from one data type to another (for example, from a string to an array of decimals).
    \item After distributing the statements with high cohesion into different functions, analyze the information dependencies between these functions, namely, what information it needs from the rest of the code and what information it provides to the rest of the code. The information it needs becomes a parameter in the function definition. The information it provides becomes a return value.
    \item Lastly, the function that was just created must be called in the appropriate place in the code. When calling the function one must pay attention to two aspects: the arguments that are being passed and whether there is a return from the function that needs to be captured. 
\end{enumerate}

% Given that the described procedure is inherently a semi-intuitive process driven by heuristics, it is understandable that as a result of the process, the resulting student code will exist on a spectrum of poorly to optimally decomposed. To make the algorithmic strategy more explicit, we outline a method of partially automating the decomposition process through the construction of a variable dependency graph.

\subsection{Scaffolded Exercises}
\label{subsec:exercises}

\begin{figure}
    \centering
    \includegraphics[width=0.4\textwidth]{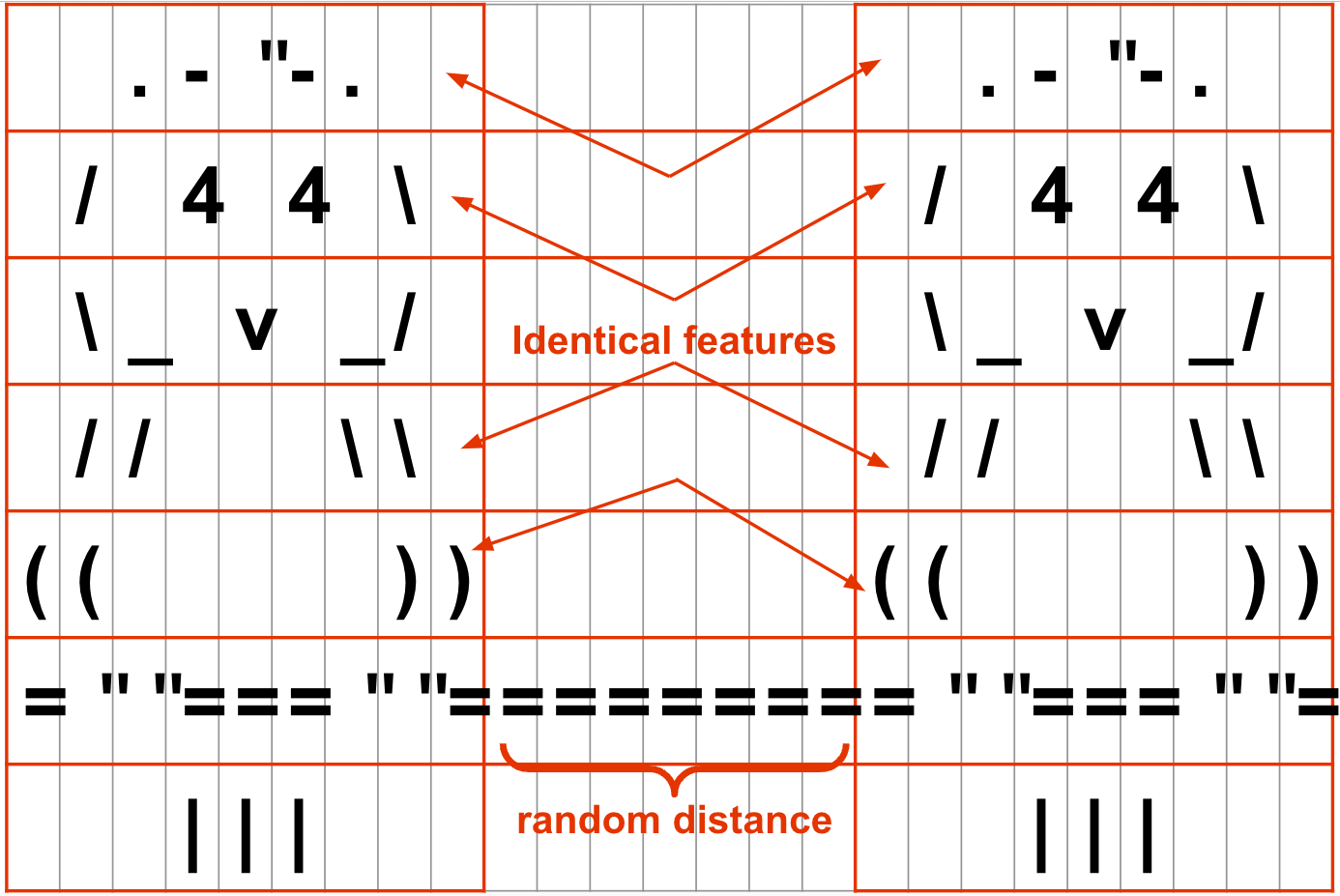}
    \caption{Exercise requiring drawing two owls (identical shape) separated by a random distance using ASCII characters. This exercise combines feature duplication (the owl parts are identical) with structural duplication (each row contains the same owl part printed twice with a random number of spaces or \texttt{=} characters for the branch in between). To increase reusability, one may chose to write only one procedure for drawing each row that takes in an owl feature and the in-between feature (either spaces or \texttt{=} characters for the branch).}
    \label{fig:owls}
\end{figure}

Structural code quality is often taught using good and bad examples of structured code alongside problem solving in worked examples. However, analyzing code is a somewhat different learning activity than actively structuring code. Problem-solving can shift the focus away from the structuring of code while following a worked example. Therefore, we propose several types of exercises that target at practice with structuring code\footnote{The exercises can be viewed at: \href{https://doi.org/10.5281/zenodo.12797876}{https://doi.org/10.5281/zenodo.12797876} }. While each type of assessment targets a slightly different feature of optimally structured code, an undecomposed set of code statements are provided in all the exercises and students are required to optimally structure them.

\subsubsection{Exercises for eliminating duplication} These kinds of programs must contain duplicated code that can be trivially identified and eliminated by simply wrapping a copy of the duplicated code with a function and calling the function repeatedly. One example is ASCII art code that draws multiple objects with repeated features using print statements with special characters, as in Figure \ref{fig:owls}.

% \begin{figure}
% \includegraphics[width=.5\linewidth]{images/owls.png}
% \caption{Exercise requiring drawing two owls (identical shape) separated by a random distance}
% \label{fig:owls}
% \end{figure}

\subsubsection{Exercises for showcasing the single-responsibility principle} These kinds of programs combine the same features or functionality in different ways. Simple exercises could involve ASCII art code that draws multiple objects with different features. The Garden Problem described in the previous section 
%and shown in Figure~\ref{fig:garden_code} 
is a more advanced assessment of this type. 

\subsubsection{Exercises for increasing reusability} These kinds of programs must contain duplicated code that is slightly different at certain points in a repeated pattern. The student must identify the pattern and utilize parameters to signal the point of difference in the pattern. Printing the verses of nursery rhyme songs (for example, the Old MacDonald song) is a simple example of this type of program because the verses are often repeated with slight variations of certain words. The owl problem shown in Figure \ref{fig:owls} is another problem where reusability can be increased by having only one method for drawing each row. Lastly, the Rubik's Cubes exercise that requires students to print how many Rubik's cubes can fit alongside each of the dimensions of a box (length, width, and height) and in total, is a more challenging example of this type of assessment. 

% This method would take in a different own feature and in-between feature (either spaces or branch) to draw.

\subsubsection{Exercises involving programming-specific tasks} These kinds of programs contain additional code that is not specified by the problem, but is needed as part of the translation of a world problem to a programming solution. For example, the Drive Times exercise takes a string as an input and needs to extract and then average the values inside that string. The pre-processing of the string to a list of values that can be averaged is a programming-specific task that is independent of the application-specific tasks.

\subsection{Supportive Automatable Visualization}
\label{subsec:visualizations}

The step-by-step procedure described in Section~\ref{subsec:novices_procedure} may be cognitively involved for students, so developing automated tools (such as a visualizer) to reinforce this process and the knowledge it leverages is also important for students' learning. The student decomposition examples in Figures~\ref{fig:garden_ex1} and ~\ref{fig:garden_ex2} suggest that students struggle to split statements with high cohesion into functions when multiple functions share data dependencies (for example, the \texttt{circle\_area} is needed to compute the information for all the outputs in the Garden problem shown in Figure~\ref{fig:garden}).

\begin{figure}
    \centering
    \includegraphics[width=0.5\textwidth]{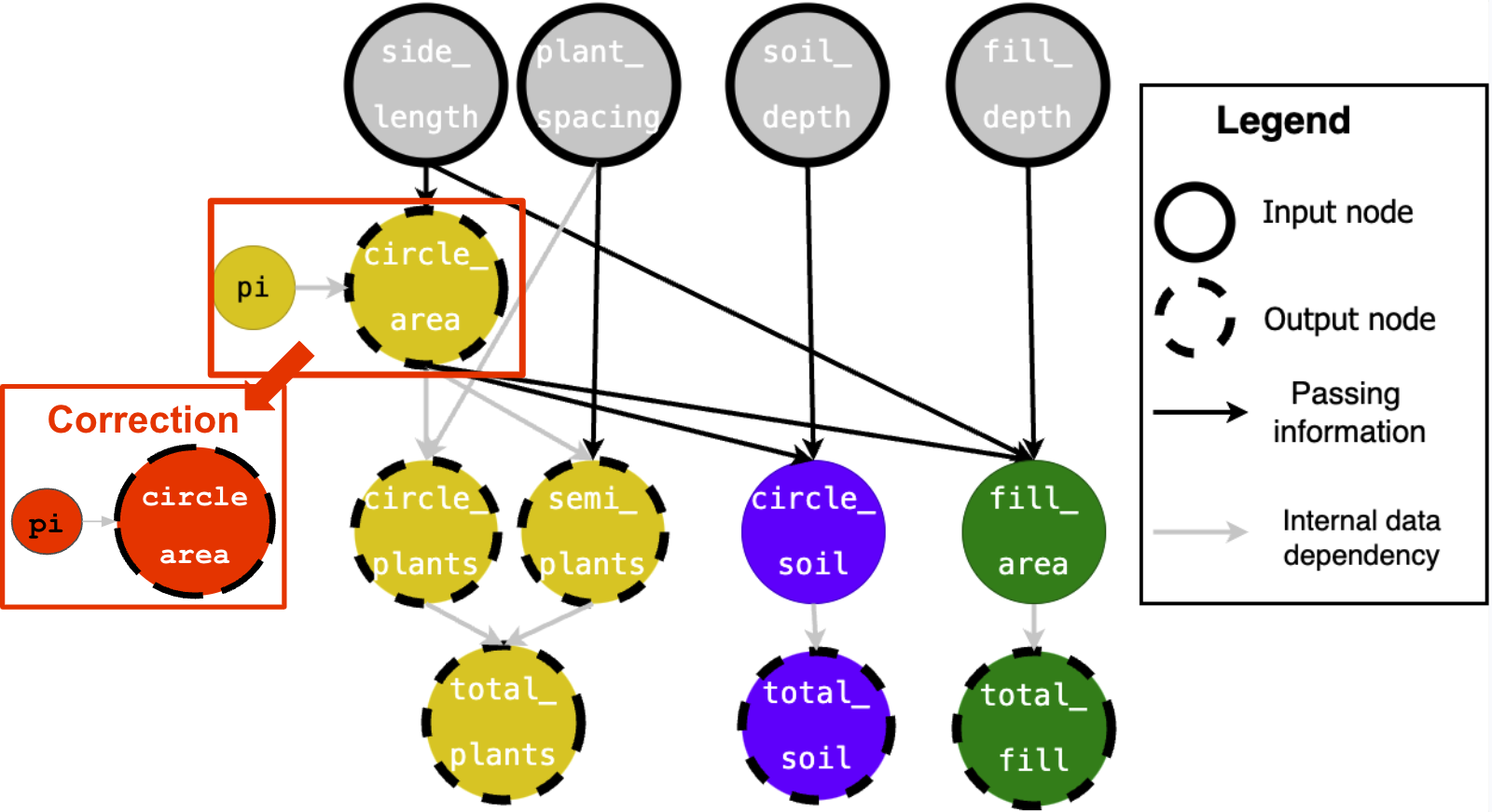}
    \caption{Depiction of the program decomposition corresponding to the student code in Figure~\ref{fig:garden_ex1} and its corresponding correction using the coloring procedure of its Data Dependency Graph (DDG) described in Section~\ref{subsec:visualizations}. Each node in the DDG represents data that is either inputted or computed and the arrows depict the dependencies between nodes. Black arrows signal that the dependency also requires the data to be passed from one procedure to another. Each color represents a separate function, namely, gray for \texttt{main}, yellow for \texttt{plants}, purple for \texttt{soil} and green for \texttt{fill\_area}. The correction corresponding to the violation of the single responsibility principle is shown in the red box on the left side of the figure. The \texttt{circle\_area} node and the nodes it depends on (such as the \texttt{pi} node) must be split away as a separate function because the information it stores is needed across three functions.}
    \label{fig:ex1_split}
\end{figure}

This process can be automatically depicted by coloring the data dependency graphs (DDG)~\cite{fenton2014software} using a backwards program split procedure~\cite{ardalani2023supporting} starting from the goals (or the outputs or prints) shown on the bottom and moving towards the sources (or the inputs) shown on the top in Figure~\ref{fig:ex1_split}. We first start with the plants goals in the lower-left corner of the DDG and move up towards the inputs coloring all the nodes with the same color, in this case yellow. Then we start from the soil goal in the lower-center corner and similarly move up towards the inputs coloring all the nodes with the same color, in this case purple. When we come across a node like \texttt{circle\_area} that is already colored, we color it and all the nodes after it with a different color from the previously used color, say red. We repeat this process until we have colored all the goals. In the end, the nodes with the same color represent a separate function\footnote{A visualization of this process can be accessed at: \href{https://youtu.be/njyxdunXIbQ}{https://youtu.be/njyxdunXIbQ}}. 

\section{Case Study}
\label{sec:case_study}

The scaffolded exercises and actionable strategy presented in this paper were piloted in the Spring 2024 semester in two sections of CS1 at a small liberal arts college. Out of the combined total of 44 students, 37 signed an informed consent to have their course data used for research. The curriculum of this CS1 course is focused on procedural Python with user-defined functions and procedural decomposition and composition being taught in the second and third week of class. The decision to introduce these topics this early was made a little over a year prior because of anecdotal evidence and supporting evidence from research about the difficulty of these topics. Despite this change, there is still room for improvement. For example, for all lab or homework assignments except one from each category, students are provided with a structure of the programs that they are required to implement to reduce the cognitive load and the variability of student designs which in turn makes grading easier (and in some cases, automatable). Despite seeing many optimal problem decomposition examples in their introductory courses, many students still fail to develop adequate program decomposition skills and thus continue to struggle in later courses.

 % The hope was also that students would learn from the problem decomposition that was done for them.

%but that was generally not the case and it is still not the case as many students continue to exhibit serious issues with decomposition in later courses. 

Students were introduced to worked examples and practiced with the type of exercises described in Section~\ref{subsec:exercises} except \textit{Tower} and \textit{Garden} which were part of a homework assignment. Up to this homework, students were introduced to the actionable strategy as it's described in Section~\ref{subsec:novices_procedure} and were not shown the data dependency graphs (DDGs) yet. After grading the homework, the instructor (who is also the main author of this paper) conducted an in-class review session to discuss the main issues in students' program decomposition. There was one other issue which was that some students kept the shared variables \texttt{side\_length}, \texttt{plant\_spacing}, \texttt{soil\_depth}, \texttt{fill\_depth} or \texttt{circle\_area} as global variables. Students were told to make all global code local in the assignment so this issue was the easiest to address. The issue with the single responsibility principle was also relatively easy to address by using the code example in Figure~\ref{fig:garden_ex2}. However, it was harder to explain why the granularity of the decomposition in Figure~\ref{fig:garden_ex1} is not optimal especially given the interconnection between different parts of the different calculations. To aid students in visualizing the longer dependency paths between different components, the instructor drew a DDG and performed the coloring similarly to how it's described in Section~\ref{subsec:visualizations}.

%drawing the data dependency graphs the instructor was able to visualize the longer dependency path between different components and give students a strategy to refine their decomposition skills.  

Additionally, there was a Parson's problem~\cite{ericson2022parsons} on the following midterm exam that required students to decompose a non-decom-posed program. This problem was one of five on a 75-minute exam and students were given some structure for this problem to get them started. During the data analysis phase, we reflected that given the simplicity of the problem with the provided structure, steps 1 and 2 of the actionable strategy presented in Section~\ref{subsec:novices_procedure} were made too easy and concluded that the assessment primarily targeted steps 3 and 4. Unsurprisingly, 34 students (92\%) correctly split the code statements into appropriate functions on the midterm assessment (which corresponds to steps 1 and 2). However, only 18 students (49\%) identified the correct parameters, returns, their correct ordering, data types, correct arguments and return capturing  (which corresponds to steps 3 and 4). We scored each feature (a total of seven) with one point and found that 5 students (14\%) got a score of two or less. 

Students were also asked to reflect on their thought process while working on this problem on the exam. The majority of the students described a systematic process similar to the actionable strategy taught in class. Only one student who scored a 6 (out of 7) described the process as \textit{trial and error until it fits}. The 5 students with scores less than 2 describe following a on-demand process driven by what was provided in the \texttt{main} function (which was the program driver that called all the auxiliary functions). One of these 5 students wrote: \textit{I first check what is displayed in main() to see what values I should get, and the sequence of getting them, so I separate the given code following the comment and return value for further usage in parameters of functions.} This student does not emphasize data dependencies as a guide for decomposition and concomitantly failed to properly split the code statements into appropriate functions. Most importantly, five students mentioned that the classroom instruction and review session were useful, for example, one student wrote \textit{[I felt] confident because of the review}.

\section{Limitations}

Despite the empirical and qualitative evidence presented in support of the proposed conceptual framework, the most obvious limitation is the lack of targeted experimental data to validate its desired qualities. These qualities include the ability to be understood by students unfamiliar with programming and code quality and the alignment of the results from the process of decomposing programs using the actionable strategy presented in Section~\ref{subsec:novices_procedure} with the expectations of educators and practitioners. 
%Verifying these qualities would allow the system to be changed so that it better aligns with the broader educational and professional community. 

%For example, further A/B testing studies would be needed to understand its short-term effects and longitudinal studies for its long-term effects. 

Moreover, close attention must be paid to the type of assessments used with the exercises we provided. For example, we concluded that the homework and exam assessment were different despite the use of similar exercises. The homework was a take-home assessment with a one week deadline. Thus, students would have been able to receive help from their peers and run the program with each change. 
%The midterm assessment was one of five problems on a 75-minutes paper exam and students were given some structure for this problem to get them started. 
The ability to run the program means that students could use the interpreter's feedback to accomplish steps 3 and 4 described in Section~\ref{subsec:novices_procedure} without necessarily understanding the mechanisms of information passing. Conversely, providing students with some structure in the exam assessment could make steps 1 and 2 too obvious given the simplicity of the programs. The difference between the assessments is supported by the lack of correlation between students' performance on the homework versus the midterm assessment. Validated assessments for program decomposition or code quality more broadly is another area of research that needs to be explored further.

Lastly, we need further studies to understand if the automatable visualization of the optimal program decomposition via coloring presented in Section~\ref{subsec:visualizations} is helpful for students' learning. A previous study~\cite{soares2020python} found that Python developers perceived data dependency graphs as less intuitive and helpful in understanding programs than control flow graphs~\cite{fenton2014software} and function call graphs~\cite{fenton2014software} which may be explained by the lack of familiarity with this type of abstraction. This may present an obstacle for the effectiveness of the proposed visualization.

\section{Conclusion}

In this paper, we present a conceptual framework that can be worked into an introductory CS curriculum to increase focus on the often neglected areas of program quality and program decomposition. One important benefit of it is that instructors can choose to use its individual components independently or together: (1) teaching program decomposition as balancing the three main concerns, (2) teaching students the actionable strategy to decompose programs with or without the use of data dependency graphs and (3) employing the scaffolded exercises targeted at program decomposition. The actionable strategy and the exercises we propose can be used to teach students how to identify decomposition patterns as well as approaches to properly decompose them. These patterns aim to promote schema formation~\cite{rist1989schema} and with the data dependency graphs as intermediate supports~\cite{arzarello1993learning} reduce the cognitive load during program decomposition. 

%Once students become more comfortable with procedural decomposition, they can use the more conceptually demanding step-by-step procedure. Together, we hope that these tools provide a comprehensive road map for teaching procedural decomposition to novices. 
%These patterns can then be used by students in the algorithmic process.   By basing the problem in a dependency graph, we hope that students will have a reduced conceptual load and will be able to more easily process the elements of the  problem that they are decomposing
% This paper describes a conceptual framework for teaching program decomposition at introductory level. 

\newpage

\bibliographystyle{ACM-Reference-Format}
\balance
\bibliography{paper}

\end{document}